\documentclass[aps, 10pt, prd, notitlepage,
twocolumn, superscriptaddress, nofootinbib, tightenlines]{revtex4-1}

\usepackage[linktocpage,breaklinks]{hyperref}
\usepackage[usenames,dvipsnames]{xcolor}
\usepackage{times}
\usepackage{newtxtext}
\usepackage{newtxmath}
\usepackage[T1]{fontenc}
\usepackage{amsmath}
\usepackage{tensor}
\usepackage[utf8]{inputenc}
\usepackage{mathrsfs}
\usepackage{bm}

\usepackage{graphicx}
\usepackage{epsfig}
\usepackage{epstopdf}

\usepackage{natbib}
\usepackage{hyperref}
\usepackage[capitalise]{cleveref}
\usepackage{multirow}

\definecolor{romared}{RGB}{142,0,28}
\hypersetup{colorlinks=true,
            citecolor=romared,
            linkcolor=romared,
            urlcolor=romared}

\newcommand{\be}{\begin{equation}}
\newcommand{\ee}{\end{equation}}

\begin{document}

\title{Thin-Shell Black Bounce}


\begin{abstract}
A recently proposed class of black-bounce solutions sourced by anisotropic fluids, whose exterior reproduces the Reissner–Nordström geometry all the way down to a minimal-area bounce, was put forward by Lessa and Olmo. In this work, we show that this solution conceals a thin-shell structure precisely at the bounce. Motivated by this observation, we construct a new class of black-bounce solutions with static, spherical symmetry featuring the emergence of a minimal-area bounce supported by a thin shell. This nontrivial topological structure is shown to arise from the degeneracy of the metric at the bounce, where the inverse metric becomes ill-defined, leading to curvature tensors with distributional discontinuities precisely at that point. As a consequence, a delta-function thin shell of energy-condition-violating matter is necessarily present at the bounce. To generate such solutions, we draw on proposals from quantum gravity that predict the existence of a fundamental minimum length scale, employing energy density profiles that encode these quantum-gravitational effects in an effective manner, replacing the point-like description of matter at minimal-length scales with a smeared one. We consider two well-motivated profiles, of Lorentzian and Gaussian type, to generate a novel class of solutions exhibiting the features of either regular black holes or wormholes, whose center conceals a thin-shell bounce. In addition, we investigate the thermodynamics of these compact objects.


\end{abstract}

\author{Leandro A. Lessa}
\email{leandrophys@gmail.com}
\affiliation{Programa de Pós-Graduação em Física, Universidade Federal do Pará, 66075-110, Belém, PA, Brazil}

\author{Renan B. Magalhães }
\email{renanbatalha@id.uff.br}
\affiliation{Instituto de Física, Universidade Federal Fluminense, Niterói, Rio de Janeiro, 24210-346, Brazil.}

\author{Gonzalo J. Olmo }
\email{gonzalo.olmo@uv.es}
\affiliation{Departamento de Física Teórica and IFIC,
Centro Mixto Universitat de Valencia - CSIC. Universitat
de Valencia, Burjassot-46100, Valencia, Spain.}

\date{{\today}}
\maketitle

\section{Introduction}
\label{sec:int}

At low energies, gravity is very well described by General Relativity (GR). However, it is widely believed that this classical description is incomplete at high energies, where quantum effects may become relevant (for instance during the early stages of the Universe~\cite{rovelli2008loop}). Moreover, spacetime singularities are \textit{unavoidable} within GR, appearing either in the past or in the future~\cite{senovilla1998singularity,senovilla20151965}. It is expected, nevertheless, that a unified quantum gravity theory---incorporating the Standard Model, gravity, and their interplay---could get rid of these pathologies~\cite{carlip2001quantum}.

Several quantum gravity frameworks provide regularization mechanisms that allows for the avoiding of singularities in the center of compact objects, for instance through the emergence of a \textit{bounce}---a local minimum of the areal radius---during the gravitational collapse~\cite{rovelli2014planck,kelly2021black}. Despite their formation, compact objects defined by a bouncing surface (from now on simply called \textit{the bounce}) at their cores have gained increasing attention in the literature~\cite{simpson2019black,lobo2021novel,mazza2021novel}. If the bounce is covered by an event horizon those objects are frequently called black bounces, while if no horizon is present these surfaces are frequently associated to wormholes and the bounce is called \textit{bounce}~\cite{visser1989traversable,visser1989traversablePRD}. 

On one hand, objects with a minimal 2-sphere core can produce distinct observable signatures\footnote{We remark that the observation of a minimal 2-sphere cannot be used to infer anything about the global properties of the spacetime~\cite{magalhaes2024echoes}.}that can be used to tell them apart from standard (singular) black holes---which has contributed to capture the attention of the Physics community on them~\cite{tsukamoto2021gravitational,ghosh2022analytical,yang2021echoes,ou2022echoes}---, on the other hand, their formation mechanism or even their (matter) source are still an open question~\cite{bueno2025dynamical}. In particular, black bounce solutions within GR were proposed to be sourced by nonlinear electrodynamics with a phantom scalar field~\cite{canate2022black,bronnikov2022field,bronnikov2022black,rodrigues2023source}, where these phantom fields are known for violating energy conditions, being a crucial element for the support of a bounce or in expanding-universe cosmological models, where they are treated as a candidate for dark energy. 

More recently, in Ref.~\cite{Lessa:2024erf}, anisotropic fluids subjected to suitable equations of state (EoS) were considered as the sources of black bounce solutions. In this formalism, the black bounce structured---encoded in the areal radius---is related to the matter sources. Thus, by specifying an EoS, black bounce solutions can be found for scalar and electromagnetic fields as well as a combination of both of them. Depending on the source, different areal radius are allowed, leading to very distinct black bounce structures. An important point on this formalism is that, provided that $p_r=-\rho$, where $\rho$ and $p_r$ are, respectively, the energy density and the radial pressure of the anisotropic fluid, the areal radius presents a local minimum if the energy density presents a maximum. 

This indicates that matter distributions with energy density everywhere finite can prevent the vanishing of the 2-sphere in favor of the emergence of a bounce. Effectively, those regularization mechanisms can be modeled by replacing the point-like localized sources with \textit{smeared} distributions with a characteristic length scale $\sqrt{\theta}$~\cite{nicolini2006noncommutative,nicolini2010noncommutative}. We recall that such formalism provides a suitable tool to investigate regular spacetime geometries. This approach has been employed in a variety of contexts, including noncommutative geometry, generalized uncertainty principle scenarios, string-inspired models, and higher-derivative gravity. 
As we shall see, these smeared distributions along with the condition $p=-\rho$ are indeed suitable for generating a bounce in the spacetime. Naively, these bounces seem to satisfy the energy conditions everywhere, leading to the question: what does keep the bounce open? In short, thin shells, where the energy conditions are unavoidably violated. Similarly as discussed in Refs.~\cite{feng2023smooth,baines2023defect} for addressing the ``defect wormholes''~\cite{klinkhamer2022defect,klinkhamer2023vacuum}.

The content of this paper is organized as follows. In Sec.~\ref{sec2}, we develop the mathematical framework for generating black bounces within GR with an anisotropic fluid with $p_r=-\rho$. We further discuss on the degeneracy of the bounce in this formalism. In Sec.~\ref{thinshell} we recall the thin-shell formalism and show that the apparent respecting of the energy conditions but degeneracy of the metric at the bounce can be a ``relic'' of a bad choice of coordinates, and that it can be understood from the thin-shell formalism as a layer violating the energy conditions. In Sec.~\ref{thinshellbounces} we consider two smeared distributions, specifically a modified Lorentzian distribution and a Gaussian distribution, to generate novel families of black bounce solutions within GR. The structure, curvature and thermal properties of these objects is studied. Finally, we summarize our results and discuss some perspectives in Sec.~\ref{con}.

\section{Equations of motion} \label{sec2}
In this section, we develop the mathematical framework necessary to understand how quantum-gravitational effects can give rise to regular compact object solutions. To this end, we first assume that general relativity, namely the Einstein--Hilbert action, is interpreted in an effective sense, with new gravitational effects arising from short-distance physics being encoded in the matter sector. This effective approach has been widely employed in the literature, primarily due to the lack of a complete theory of quantum gravity. In fact, progress in approaches such as loop quantum gravity and loop quantum cosmology suggests that an effective metric description offers a reliable approximation across a wide range of scenarios.
 (see, e.g., Refs.\cite{PhysRevD.81.084027,ashtekar2015generalized}. Accordingly, we consider the following action:
\begin{equation}\label{ac}
    S = \frac{1}{16\pi G}\int d^{4}x\sqrt{-g}\bigg[R  +\mathcal{L}_m \bigg],
\end{equation}
where the $G$ represents Newton’s gravitational constant, and $\mathcal{L}_m$ is the matter sector. The equations can be written as
\begin{equation}\label{eqq1}
    G_{\mu\nu} = 8 \pi G T_{\mu\nu},
\end{equation}
 where  the $G_{\mu\nu} = R_{\mu\nu}+\frac{1}{2} g_{\mu\nu}R$  is the Einstein tensor and the $T_{\mu\nu}$ is the matter stress-energy tensor. 

 For spherically symmetric and static configurations, and with the aim of being as general as possible in order to explore the widest range of geometries, we can parametrize the line element as
\begin{equation} \label{1}
 ds^{2} = - A(x)dt^2 + B(x) dx^2 + r(x)^2d\Omega^2
 \end{equation}
where $d\Omega^2=d\theta^2 + \sin^2\theta d\phi^2$ is the element of solid angle on a $2$-sphere, whose areal radius is given by $S=4\pi r(x)^2$ and must be everywhere gratear tahn zero. It is important to emphasize that the radial coordinate is $x$, defined over the interval $(-\infty, \infty)$, and that $r(x)$ is the so-called areal function. In some cases, as we shall mention later, it is indeed possible to use this function as a radial coordinate. However, when it is not well defined, this choice becomes inappropriate. In fact, as pointed out in Ref.~\cite{LimaJunior:2025uyj}, this function can be classified into two categories: monotonic and non-monotonic. The focus of this work lies precisely on the latter case, namely when $dr/dx = 0$ at some location, in which case it is not possible to use $r$ as a valid coordinate over the entire domain. Consequently, a non-monotonic areal function may encode important geometric information that cannot be trivialized by a simple coordinate transformation.
 
 Since the main focus of this work is to further investigate the properties of geometries in which the area of the two-spheres may exhibit a nonvanishing lower bound (i.e., a nonmonotonic behavior of the areal function $r$), the next step is to identify matter sources capable of supporting such geometries, while general relativity is assumed to be the effective geometric theory governing the system
. As shown in Ref.~\cite{Lessa:2024erf}, this can be achieved, for instance, within the framework of nonlinear electrodynamics. In the present work, however, we assume that the matter sector effectively arises from quantum-gravitational corrections at short-distance scales, where the pointlike mass structure is replaced by a smeared distribution. Adopting an effective framework, we encode the quantum gravitational effects in the matter sector through an anisotropic fluid description.
 Consequently, the stress--energy tensor is taken to be that of an anisotropic fluid, given by


\begin{equation} \label{flui}
    T_{\mu\nu} =(\rho+p_{t})l_{\mu}l_{\nu} + p_t g_{\mu\nu} + (p_r-p_t)v_{\mu}v_{\nu}
    \end{equation}
where $\rho$, $p_r$, and $p_t$ represent the energy density, the radial pressure, and the tangential pressure of the fuid, respectively. The vector $l^{\mu}$ represents the four-velocity of the fluid, normalized as $g_{\mu\nu} l^\mu l^\nu=-1$, while $v_{\mu}$ denotes a space-like unit vector normalized as $g_{\mu\nu} v^\mu v^\nu=+1$.  According to the line element~\eqref{1}, the vectors $u^\mu$ and $v^\mu$ take the form
\begin{eqnarray}
    l^\mu=(\frac{1}{\sqrt{A(x)}},0,0,0) \ , \nonumber \\
    v^\mu=(0,\frac{1}{\sqrt{B(x)}},0,0) \ .
\end{eqnarray}
Additionally, raising one index of the stress-energy tensor with the metric, we see that it becomes diagonal and with components ${T^\mu}_\nu=  \frac{1}{8\pi G}\text{Diag}(-\rho,p_r,p_t,p_t)$.

Within this simple setup, i.e.,  general relativity supplemented by effective quantum corrections in the matter sector and assuming a static, spherically symmetric geometry, we now investigate how these ingredients interact through the equations of motion. Thus, the field equations (\ref{eqq1}) evaluated on the line element (\ref{1}) become
\begin{equation}\label{eq1}
\frac{B' r '}{B^2 r}-\frac{2 r ''}{B r}-\frac{r '^2}{B r ^2}+\frac{1}{r ^2} =  \rho, 
\end{equation}
\begin{equation}\label{eq2}
   \frac{A' r '}{A B r }+\frac{r '^2}{B r ^2}-\frac{1}{r ^2} = p_r
\end{equation}
and
\begin{equation}\label{eq3}
  \frac{A''}{2 A B}-\frac{A' B'}{4 A B^2}+\frac{A' r '}{2 A B r}-\frac{A'^2}{4 A^2 B}-\frac{B' r '}{2 B^2 r}+\frac{r ''}{Br }=p_t.
\end{equation}
On the other hand, from the conservation of the stress-energy  tensor, $\nabla_{\mu}T^{\mu}{}_{\nu}=0$ one obtains
\begin{equation} \label{conse}
  \frac{A'}{4 A}\bigg( \frac{\rho+p_r}{p_t-p_r} \bigg)+\frac{p_r '}{2(p_t-p_r)}=    \frac{r '}{r } \ .
\end{equation}
The above set of four equations involves six unknown functions, namely, $\left\{A, B, r, \rho, p_r, p_t \right\}$, which demands some additional inputs in order to solve the system. Since our focus here is to understand the mechanisms that determine the dependence of $r(x)$ on the matter sources, we will impose some restrictions on the matter sector to facilitate the analysis. To motivate the constraints we will use, let us first add Eqs.(\ref{eq1}) and (\ref{eq2}) to obtain
\begin{equation} \label{AB0}
    \frac{1}{B}\bigg[ \bigg(\frac{1}{AB}\frac{d(AB)}{dx}\bigg)\frac{ r '}{r} -\frac{2 r ''}{ r} \bigg]=\rho+p_r.
\end{equation}

Given the expressions (\ref{conse}) and (\ref{AB0}), we see that important progress can be achieved by considering matter sources for which the combination $\rho+p_r$ vanishes. This vacuum-like condition, which is invariant under boosts in the radial direction, has been widely employed in the literature to construct static, spherically symmetric regular black hole solutions with a de Sitter core~\cite{gliner1966algebraic,dymnikova1992vacuum,bronnikov2003nonsingular,Zaslavskii:2025oli}.
Restricting our attention to such kind of sources, we find that (\ref{AB0}) becomes 
\begin{equation} \label{AB}
    \frac{1}{B}\bigg[ \bigg(\frac{1}{AB}\frac{d(AB)}{dx}\bigg)\frac{ r '}{r} -\frac{2 r ''}{ r} \bigg]=0,
\end{equation}
which can be integrated once to yield 
\begin{equation}\label{rel}
    B = \frac{r '^2}{A} \ ,
\end{equation}
where an irrelevant integration constant has been absorbed in a redefinition of the coordinate $x$. Thus the line element is given by
\begin{equation} \label{Metricc}
 ds^{2} = - A(x)dt^2 + \frac{r(x) '^2}{A(x)} dx^2 + r(x)^2d\Omega^2.
 \end{equation}
That is, it is easy to see that it can be written as
\begin{equation} \label{1m}
 ds^{2} = - A(r)dt^2 +  \frac{dr^2}{A(r)} + r^2d\Omega^2 \ ,
 \end{equation}
which is a very standard form of writing the line element when $r$ is used as the radial coordinate. At this point, it becomes clear that when the areal function has a minimum at a given point, the metric is degenerate at that point in the original radial coordinate x. However, when we use the radial coordinate r, this information is hidden. For this reason, we will explore the effects introduced by a metric with a non-monotonic areal function.

The form of the function $A(r)$ can then be obtained from Eq. (\ref{eq2}) by assuming that $p_r=-\rho$ can be written as a function of $r$, which leads to 
\begin{equation}\label{eq:AofSigma}
    r A_{r} + A = 1 - \rho r^2 \ ,
\end{equation}
where we are denoting $A_{r}= \frac{\partial A}{\partial r}$. This equation can be further simplified with the ansatz $A=1-2M(r)/r$, leading to $\frac{\partial M}{\partial r}=\rho r^2/2$, which is the usual Newtonian expression for the mass function and can be directly integrated if a function $\rho=\rho(r)$ is given. This is what we discuss next.

We can now turn our attention to the conservation equation  (\ref{conse}), which under the assumption $\rho+p_r=0$ turns into 
\begin{equation} \label{conse1}
 \frac{r '}{r }=-\frac{\rho'}{2(p_t+\rho)}
\end{equation}
To proceed, we assume that the matter source also satisfies a linear equation of state described by $p_t = \omega \rho$, where $\omega$ is a constant. By specifying this equation of state, the Eq. (\ref{conse1}) can be integrated by quadratures, yielding a relation of the form 
\begin{equation}\label{energ}
r(x)=r_0\bigg(\frac{\rho(x)}{\rho_0}\bigg)^{-\frac{1}{2(1+\omega)}},
\end{equation}
where $\rho_0$ represents the maximum density at the bounce
. Note that if $\rho(x)/\rho_0 > 0$ for all $x$ in its domain and $\omega > -1$, then if the function $\rho(x)$ has a maximum, the areal function will have a minimum at the same point.
We can now use this energy density expressed as a function of $r(x)$ to solve the Eq. \eqref{eq:AofSigma}, where $r$ is treated as the radial coordinate in this case. After integrating, we find that
\begin{equation}\label{eq:ASigma}
    A(r)=\left\{\begin{array}{lr} 
    1-\dfrac{2\tilde{m}}{r(x)}-\dfrac{\rho_0 r_0^2}{(1-2\omega)}\left(\dfrac{r_0}{r(x)}\right)^{2\omega} & \text{ if } \omega\neq 1/2 \\
    1-\dfrac{2\tilde{m}}{r(x)}-\dfrac{\rho_0 r_0^3}{r(x)}\ln\left(\dfrac{r(x)}{r_0}\right) & \text{ if } \omega= 1/2\end{array}\right. \ ,
\end{equation}
where $\tilde{m}$ is an integration constant that represents the asymptotic ADM mass. 

From the equations of state used to derive Eqs.~\eqref{energ}~\eqref{eq:ASigma}, namely $p_r = -\rho$ and $p_t = \omega \rho$, one can show that this fluid seems, in fact, to satisfy the energy conditions. Indeed, the null energy condition (NEC), given by the inequalities $\rho + p_r \ge 0$ and $\rho + p_t \ge 0$, is satisfied provided that $\rho \ge 0$ and $\omega \ge -1$. Consequently, the weak energy condition (WEC), defined by $\rho \ge 0$, $\rho + p_r \ge 0$, and $\rho + p_t \ge 0$, is also satisfied. The strong energy condition (SEC), written as $\rho + p_r + 2 p_t \ge 0$, holds if $\omega \ge 0$ for $\rho \ge 0$. It remains to examine the dominant energy condition (DEC), which requires $\rho + |p_r| \ge 0$ and $\rho + |p_t| \ge 0$. These inequalities imply the conditions $\rho + p_r \ge 0$, $\rho + p_t \ge 0$, $\rho - p_r \ge 0$, and $\rho - p_t \ge 0$, all of which hold for positive energy density provided that $\omega \le 1$. Remarkably, the case $\omega = 1$ is particularly interesting, as it not only satisfies all the energy conditions but also meets the requirement for generating a non-monotonic areal function with a bounce located at the local maximum of the energy density $\rho$. For this reason, we will focus on this case of interest throughout the remainder of this work.

Having established the mathematical framework to be adopted throughout this work, we next investigate the gravitational configuration described by the metric~\eqref{Metricc} with \eqref{eq:ASigma}. Furthermore, we assume that the energy density sourcing this object exhibits a maximum at $x = x_{\text{bounce}}$, thereby generating a bounce structure at this location \eqref{energ}, which implies $r'(x_{\text{}}) = 0$ along with the flare-out condition $r''(x_{\text{bounce}}) > 0$. However, as demonstrated above, the chosen anisotropic fluid does not violate the energy conditions; thus, in principle, it should not be capable of generating an object with a non-trivial topology at its core\footnote{This expectation is corroborated by the topological censorship theorem, which shows that in a globally hyperbolic, asymptotically flat spacetime satisfying the null energy condition, no causal curve can probe a non-trivial spacetime topology \cite{FriedmanSchleichWitt1993}. This result was later shown to hold generically for both static and dynamical traversable wormhole bounces, where violation of the null energy condition is a necessary geometric requirement \cite{HochbergVisser1997,HochbergVisser1998}.}. This raises a fundamental question: what keeps this bounce open? The answer to this question will be addressed in the following.

\section{THIN-SHELL FORMALISM}
\label{thinshell}
This section addresses the question posed at the end of the previous one, showing that the line element \eqref{Metricc} conceals a thin-shell structure precisely at the bounce. The first indication of this behavior emerges from the metric itself, which exhibits a degeneracy at the bounce; that is, the metric is smooth, but the inverse metric is not everywhere continuous, being undefined at the bounce itself, as previously noted in a similar context in Ref.~\cite{feng2023smooth,baines2023defect}. However, to clearly illustrate that we can generate compact objects with a bounce supported by a thin shell, assuming only a certain non-exotic anisotropic fluid that models a non-monotonic areal function, we must analyze the extrinsic curvature at the bounce. We will show below that, for this class of solutions \eqref{Metricc}, a discontinuity occurs in the extrinsic curvature tensor at the bounce, thereby demonstrating the presence of a hidden thin shell. Before proceeding, we briefly review some important concepts and definitions within the thin-shell formalism, following the standard treatments of Refs. \cite{Poisson2004,barrabes1997singular}.

The thin-shell formalism provides a systematic framework for constructing spacetimes 
composed of two distinct geometric sectors. Let $\mathcal{M}^{\pm}$ be two smooth 
Lorentzian manifolds, each endowed with a metric tensor $g^{\pm}_{\mu\nu}$. By 
excising each manifold along a hypersurface $\Sigma^{\pm}$\footnote{This codimension-one submanifolds can be timelike, spacelike, or null. However, throughout this work, we restrict our attention exclusively to the first two types of hypersurfaces.}, one obtains 
two bounded regions that are subsequently identified at their respective boundaries, 
yielding a single manifold
\begin{equation}
    \mathcal{M} = \mathcal{M}^{-} \cup \mathcal{M}^{+},
\end{equation}
whose two sectors are separated by a thin hypersurface
\begin{equation}
    \Sigma = \Sigma^{\pm} = \mathcal{M}^{-} \cap \mathcal{M}^{+},
\end{equation}
across which the geometries governed by $g^{-}_{\mu\nu}$ and $g^{+}_{\mu\nu}$ are 
matched. In general, both geometric and matter quantities may exhibit 
discontinuities across $\Sigma$ , so that a distributional framework 
becomes necessary; specifically, tensorial distributions must replace ordinary 
tensorial functions in order to properly account for such singular behavior.

Within this setting, all relevant geometric and matter quantities are required to 
satisfy at $\Sigma$ the so-called \textit{junction conditions} \cite{israel1966singular,visser1995lorentzian}. The 
first of these conditions demands the continuity of the induced metric across the 
hypersurface,
\begin{equation}
    \left[h_{\mu\nu}\right] \equiv \left. g^{+}_{\mu\nu} \right|_{\Sigma} 
    - \left. g^{-}_{\mu\nu} \right|_{\Sigma} = 0,
    \label{eq:first_junction}
\end{equation}
which ensures that $\Sigma$ is a well-defined geometric object shared by both 
manifolds. The second junction condition, in turn, governs the 
discontinuity of the extrinsic curvature $K_{\mu\nu}$ across $\Sigma$ and relates 
it to the energy-momentum content of the thin shell. Denoting the jump of any 
quantity $\mathcal{Q}$ across $\Sigma$ as 
$\left[\mathcal{Q}\right] \equiv \mathcal{Q}^{+}|_{\Sigma} - 
\mathcal{Q}^{-}|_{\Sigma}$, this condition reads
\begin{equation}
    \left[K_{\mu\nu}\right] - h_{\mu\nu}\left[K\right] = -\, 8 \pi GS_{\mu\nu},
    \label{eq:second_junction}
\end{equation}
where $K \equiv h^{\mu\nu}K_{\mu\nu}$ is the trace of the extrinsic curvature, 
$h_{\mu\nu}$ is the induced metric on $\Sigma$, and $S_{\mu\nu}$ is the 
surface energy-momentum tensor of the thin shell, which encodes the 
matter content localized on the hypersurface. In the particular case where 
$\left[K_{\mu\nu}\right] = 0$, the hypersurface $\Sigma$ carries no surface 
energy-momentum tensor, and the matching is said to be \textit{smooth}.

Let us now turn to the computation of the extrinsic curvature associated with the 
constant-$x$ hypersurfaces for~\eqref{Metricc}. Working in the coordinate system 
$(t, x, \theta, \varphi)$, the unit normal $n^{\mu}$ to these timelike hypersurfaces is 
constructed as
\begin{equation}
    \qquad  
    n^{\mu} = \frac{\sqrt{\epsilon A(x)}}{|r'(x)|}\,(0,\,1,\,0,\,0),
    \label{eq:unit_normal}
\end{equation}
where $n^{\mu}n_{\mu}=\epsilon$, i.e., for $\epsilon = -1$, we obtain a spacelike hypersurface, whereas $\epsilon = 1$ corresponds to a timelike one. For a line element with a wormhole interpretation ($\epsilon = 1$), the normal vector is deliberately oriented to always point from one asymptotic region into the other, i.e., in the direction of increasing $x$. This choice ensures a consistent notion of the ``outward'' direction across the bounce.


With the unit normal at hand, the extrinsic curvature of the constant-$x$ 
hypersurfaces is defined via the Lie derivative of the metric along $n^{\mu}$,
\begin{equation}
    K_{\mu\nu} = -\nabla_{(\mu}n_{\nu)} = \frac{1}{2}\,\mathcal{L}_{n}\,g_{\mu\nu},
    \label{eq:extrinsic_def}
\end{equation}
which, upon expanding the Lie derivative explicitly, yields
\begin{equation}
    K_{\mu\nu} = \frac{1}{2}\left[
        n^{\lambda}\,\partial_{\lambda}\,g_{\mu\nu} 
        - \left(\partial_{\mu}n_{\lambda}\right)g^{\lambda}{}_{\nu} 
        - \left(\partial_{\nu}n_{\lambda}\right)g^{\lambda}{}_{\mu}
    \right].
    \label{eq:extrinsic_expanded}
\end{equation}
%
%
%

The covariant components of the extrinsic curvature tensor $K_{\mu\nu}$ associated 
with the constant-$x$ hypersurfaces can be assembled into matrix form as
\begin{equation}
K_{\mu\nu} = \text{sgn}\left(r'\right)
\begin{pmatrix}
K_1& 0 & 0 & 0 \\[10pt]
0 & 0& 0 & 0 \\[10pt]
0 & 0 & K_2 & 0 \\[10pt]
0 & 0 & 0 & K_2\sin^{2}\!\theta
\end{pmatrix},
\label{eq:extrinsic_matrix}
\end{equation}
where the sign function $\mathrm{sgn}(r') \equiv r'/|r'|$ explicitly encodes the orientation of $r'(x)$ 
on each side of the bounce and the expressions for the components of the extrinsic curvature tensor are given by
\begin{eqnarray}\label{k1}
    K_1 = -\frac{\sqrt{\epsilon A} A'}{2 r'}
\end{eqnarray}
\begin{eqnarray}\label{k2}
    K_2 = \sqrt{\epsilon A} r
\end{eqnarray}
Therefore, we can generally conclude that if the functions $K_1$ and $K_2$ remain finite and nonvanishing at the bounce, the presence of the step-function implies that the second fundamental form $K_{\mu\nu}$ will exhibit a discontinuity precisely at the bounce $x_{bounce}$, a feature characteristic of thin-shell solutions.

In addition to this evidence in favor of a thin-shell structure at the bounce of the spacetime under consideration, we can investigate the expansion scalar $\Theta$ for a radial geodesic congruence. These geodesics satisfy the following equation:
\begin{equation} \label{geod}
    g_{\mu\nu} \dot{x}^{\mu}\dot{x}^{\nu} = k
\end{equation}
where $\dot{x}^{\mu} = \mathrm{d}x^{\mu}/\mathrm{d}\lambda$ denotes the tangent vector to the particle worldline, parametrized by an affine parameter $\lambda$. Moreover, $k=-1$ corresponds to massive timelike particles, while $k=0$ characterizes null-like particles.
Note that the spacetime~\eqref{1} is invariant under time reversal and rotations about the azimuthal angle. Consequently, there exist two Killing vectors associated with these symmetries. As a result, the quantities $E = -A\,\dot{t}$ and $L = x^{2}\,\dot{\phi}$ are conserved, , where we assume trajectories confined to the equatorial plane $\theta = \pi/2$. 

This scalar is defined as
\begin{equation}\label{expansion}
    \Theta = \nabla_{\mu}u^{\mu} = \frac{1}{\sqrt{-g}}\partial_{\mu}(\sqrt{-g} u^{\mu}),
\end{equation}
where $u^{\mu}$ is the velocity field of a family of massive or massless particles moving along a congruence of radial or non-radial ($L\neq0$) geodesics. Assuming the metric~\eqref{Metricc} possesses a Killing vector associated with time-reversal and rotations about the azimuthal
angle symmetry, the energy $E = -A \dot{t}$ and $L = x^{2}\,\dot{\phi}$ are conserved, yielding
\begin{equation} \label{u}
    u^{\mu} = \left(-\frac{E}{A(x)}, -\frac{\sqrt{A(x)\bigg(k+\frac{E^2}{A(x)}-\frac{L^2}{r(x)^2}\bigg)}}{|r'(x)|},0,\frac{L}{r(x)^2}\right),
\end{equation}
where we have chosen the ingoing geodesics. Substituting Eq.~\eqref{u} into the Eq.~\eqref{expansion}, we obtain
\begin{eqnarray} \label{exp}
        \Theta = \frac{\Theta_0}{\text{sgn}\left(r'\right)},
\end{eqnarray}
where
\begin{eqnarray}
    \Theta_0 = \frac{r A' \left(L^2-k r^2\right)+2 r' \left(A L^2-2 r^2 \left(A k+E^2\right)\right)}{2 r^3 r' \sqrt{A \left(k-\frac{L^2}{r^2}\right)+E^2}}.
\end{eqnarray}
Once again, a discontinuity emerges at the bounce for finite $ \Theta_0$. In this case, we observe that the expansion scalar changes sign as the congruence passes through the bounce.

In light of these results, we can conclude that the aforementioned discontinuities reveal the presence of a thin-shell structure within the object described by the metric~\eqref{Metricc}. Furthermore, such properties are commonly utilized in spherically symmetric compact objects to demonstrate the violation of the energy conditions. Indeed, the Einstein field equations, when applied to the hypersurface joining the bulk spacetimes, yield the Lanczos equations~\eqref{eq:second_junction}. Due to the spherical symmetry of the setup, the jump in the extrinsic curvature tensor can be written as $[K^{i}{}_{j}] = \text{diag}([K^{t}{}_{t}], [K^{\theta}{}_{\theta}], [K^{\phi}{}_{\phi}])$, which allows the surface stress-energy tensor to reduce to $S^{i}{}_{j} = \text{diag}(-\sigma, P, P)$, where $\sigma$ is the surface energy density and $P$ is the surface pressure. Consequently, the Lanczos equations imply
\begin{eqnarray}
    \sigma =- \frac{1}{4\pi G}[K^{\theta}{}_{\theta}]
\end{eqnarray}
\begin{eqnarray}
    P = \frac{1}{8\pi G}([K^{\theta}{}_{\theta}]+[K^{t}{}_{t}]),
\end{eqnarray}
where $[K^{t}{}_{t}] = K^{t}{}_{t}|_{x=x_{bounce^{+}}}-K^{t}{}_{t}|_{x=x_{bounce^{-}}}$ and $[K^{\theta}{}_{\theta}] = K^{\theta}{}_{\theta}|_{x=x_{bounce^{+}}}-K^{\theta}{}_{\theta}|_{x=x_{bounce^{-}}}$.

Using the computed extrinsic curvature~\eqref{eq:extrinsic_matrix}, we now evaluate the surface stresses:
\begin{eqnarray}\label{ener}
       \sigma = - \frac{\sqrt{\epsilon A}}{2\pi G r}\bigg|_{x=x_{bounce}}, 
\end{eqnarray}
\begin{eqnarray}\label{pre}
       P =  \frac{1}{8\pi G }\bigg( \frac{2\sqrt{\epsilon A}}{r}+\frac{A'}{\sqrt{\epsilon A}r'} \bigg)\bigg|_{x=x_{bounce}}.
\end{eqnarray}
Note that the surface energy density is strictly negative, thereby implying a violation of the energy conditions within this thin-shell framework. Furthermore, it is important to point out that the behavior of the solution $A$ at the bounce, namely, whether this metric function is positive or negative, determines the causal nature of the thin-shell hypersurface, ensuring that the quantities~\eqref{ener} and~\eqref{pre} remain real. For instance, if $A < 0$ at the bounce, then $\epsilon = -1$, which implies that the hypersurface $\Sigma$ is spacelike; conversely, if $A > 0$, then $\epsilon = 1$, meaning that the hypersurface $\Sigma$ is timelike.

\section{Thin-Shell Black Bounce on Smallest Scales}
\label{thinshellbounces}
In the previous section, we demonstrated that the spacetime~\eqref{Metricc} conceals a bounce supported by a thin shell surface stress-energy tensor that violates the energy conditions. This spacetime is sourced by a matter content that exhibits a remarkably simple configuration: $\rho + p_r = 0$, $p_t = \rho$, with $\rho$ possessing a maximum. This latter condition is essential, as it guarantees that the areal function is non-monotonic, i.e., it possesses a minimum~\eqref{energ}. In other words, assuming an energy density profile that exhibits a smoothly differentiable peak at the center governed by the aforementioned equations of state, a thin-shell bounce structure will inevitably form at this location.

In the following, we motivate the formation of this thin-shell structure via quantum-gravitational effects within an effective framework of quantum gravity models. Specifically, we utilize energy density profiles that replace a pointlike Dirac delta distribution with a smeared distribution, i.e., a Gaussian-like profile instead of a singularity at the origin. This is precisely what is required to ensure that the areal function possesses a minimum. Consequently, we justify the discontinuity in the extrinsic curvature and the sign change of the expansion scalar as physical consequences of quantum spacetime fluctuations.

\subsection{Modified Lorentzian Distribution}
On this section we begin by considering a like-Lorentzian distribution given by \cite{nicolini2019quantum,nozari2008hawking,nicolini2012nonlocal,rizzo2006noncommutative,sharif2018lorentz,araujo2024effects}
\begin{equation} \label{dens}
    \rho(x) =  \frac{\tilde{M}}{(x^2+l_0^2)^n}.
\end{equation} where $\tilde{M}$ is related to the mass parameter diﬀused throughout the
region of linear size $l_0$ and $n$ is a positive constant. The quantity $l_0$ is the fundamental minimal length predicted by different quantum gravity theories. For example, in noncommutative gravity within the coherent state formalism, we identify the minimal length $l_0$ with the noncommutativity parameter $\tilde{\Theta}$, and in this case we have $n=2$. On the other hand, in stringy corrections to black hole spacetimes emerging from string T-duality, $l_0$ is identified with the zero-point length of spacetime $l_0$, with $n = 5/2$. 

Now we can find the relation between the ADM mass and the smeared mass distribution through 
\begin{equation}
   M = \int_{0}^{x} 4\pi \rho(x) x^2dx = \frac{4 \pi  \tilde{M}   \, _2F_1\left(1,\frac{5}{2}-n;\frac{5}{2};-\frac{x^2}{l_0^2}\right)x^3}{3 l_0^2\left(l_0^2+x^2\right)^{n-1}},
\end{equation}
where $_2F_1$ is a hypergeometric function.  Therefore, to ensure that in the limit $l_0 \to 0$ we obtain $M = \tilde{m}$, we must impose that:
\begin{equation}
    \tilde{M} = \frac{l_0^{2n-3}\Gamma (n)\tilde{m}}{\pi^{3/2}\Gamma (n-3/2)} .
\end{equation}
In order to ensure that the above quantity remains positive, we require $n \in (0, 1/2) \cup (3/2, \infty)$. Furthermore, it is clear from this expression that in the limit $l_0 \to 0$ the quantum-gravitational matter source vanishes, and Einstein gravity is recovered.


For the reasons discussed previously, we set $\omega=1$. Substituting this distribution~\eqref{dens} into Eq. (\ref{energ}), we find that
\begin{equation} \label{rlorentz}
    r(x)= r_0 \left(x^2+l_0^2 \right)^{\frac{n}{4}}
\end{equation}
where we set $\rho_0=\tilde{M}$ and $r_0$ is an irrelevant integration constant that can be set to unity without loss of generality. Note that the size of the bounce at $x=0$ is given by $r(0)= l_0^{\frac{n}{2}}$, that is, the size of the bounce is of the order of the minimal length predicted by the quantum theory of gravity. Furthermore, the standard areal radius is modified in limit $l_0=0$, since $S=4\pi x^{n}$ for $n\neq2$. Near the origin the areal function can be expanded as
\begin{equation}
    r(x)\approx l_0^{\frac{n}{2}}\bigg(1 + \frac{n}{4}\frac{x^2}{l_0^2}+\frac{(n-4)n}{32}\frac{x^4}{l_0^4} + \ldots \bigg).
\end{equation}

To close the system, we can determine the metric component $A(x)$ from Eq.~\eqref{eq:ASigma} with $\omega=1$, which yields the line element given by
\begin{align} \nonumber \label{n22}
   & ds^2 = -\bigg(1-\frac{2\tilde{m}}{( x^2+l_0^2)^\frac{n}{4}} +\frac{\alpha_nl_0^{2n-3}\tilde{m}}{( x^2+l_0^2)^\frac{n}{2}}\bigg)dt^2  \\
    &+ \frac{(n^2/4)dx^2}{  x^{-2} \bigg(x^2+l_0^2\bigg)^{\frac{4-n}{2}}\bigg(1-\frac{2\tilde{m}}{( x^2+l_0^2)^\frac{n}{4}} +\frac{\alpha_n l_0^{2n-3}\tilde{m}}{( x^2+l_0^2)^\frac{n}{2}} \bigg)}+\\ \nonumber &(x^2+l_0^2)^{\frac{n}{2}}d\Omega^2.
\end{align}
where $\alpha_n = \frac{\Gamma(n)}{\pi^{3/2}\Gamma(n-\frac{3}{2})}$. Thus, to ensure the convergence of the Gamma function, we must require $n > 3/2$. Thus, we further restrict the range of this parameter to $n \in (3/2, \infty)$, ensuring the positivity of the ADM mass. Moreover, note that in the limit $l_0 \to 0$, we recover the Schwarzschild solution (if $n> 3/2$). To see this, it is necessary to perform the following coordinate transformation: $x^{\frac{n}{2}} = r$. This transformation leads precisely to the line element~\eqref{1m}, where the metric function $A(r)$ can be written as
\begin{equation}\label{1mlorentz}
    A(r) = 1 - \frac{2\tilde{m}}{r} + \frac{\alpha_n l_0^{2n-3}\tilde{m}}{r^2},
\end{equation}
namely, a modified Reissner-Nordstr\"om-like solution.

Remarkably, we observe that the solution~\eqref{n22} closely resembles the black-bounce solution found by Simpson and Visser \cite{Simpson:2018tsi}\footnote{In fact, for $n=2$, the solution~\eqref{n22} closely resembles the charged black-bounce solution \cite{Franzin:2021vnj}; however, in this case, we only have mass parameters, with no charge involved.}. This compact object represents a theoretical spacetime geometry featuring a regular bounce structure that, depending on the value of a specific parameter, either acts as the minimal radius of a wormhole or remains hidden behind a black hole horizon. However, unlike the Simpson-Visser case, the solution~\eqref{n2} introduces a novel factor, $x^{2} (x^2+l_0^2)^{\frac{n-4}{2}}$, in the radial component of the metric. Its origin stems precisely from $r'(x)^2$. This crucial difference is responsible for inducing the discontinuities in the extrinsic curvature tensor. For this reason, we can conclude that while the solution~\eqref{n2} describes a black bounce, the bounce located at the origin is actually supported by a thin shell. 

\begin{figure}[!ht] 
\centering
\includegraphics[height=5cm]{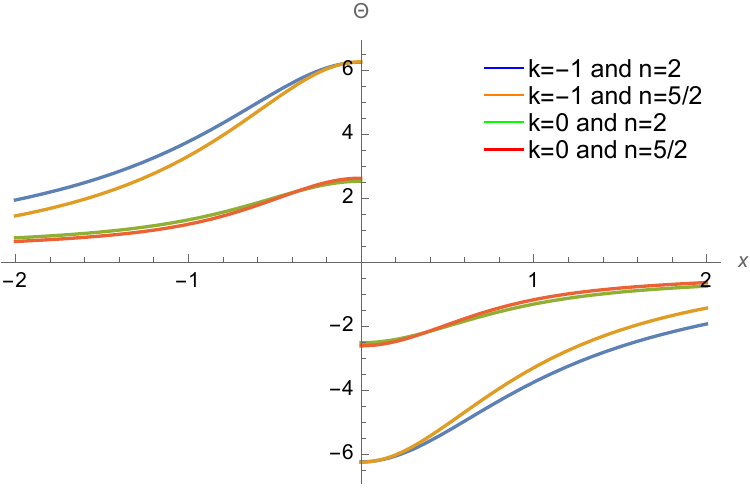}
\caption{ The representation of the expansion scalar $\Theta$~\eqref{exp} from the line element (\ref{n22}) as a function of the radial coordinate $x$. We assume that $\tilde{m}=10$, $l_0=1$, $E=1$, and $L=1$. The bounce is located at the origin}  \label{expan}
\end{figure}

We can explicitly demonstrate this by computing the extrinsic curvature tensor for this solution. Substituting Eq.~\eqref{n22} into Eqs.~\eqref{k1} and~\eqref{k2} yields 
\begin{align}
    K_1(0) = \frac{\tilde{m}   l_0^{-n-3} \left(l_0^{\frac{3 n}{2}} \Gamma (n)-\pi ^{3/2} l_0^3 \Gamma \left(n-\frac{3}{2}\right)\right)}{\pi ^{9/4} \Gamma \left(n-\frac{3}{2}\right)}\\\nonumber
   \times \sqrt{\pi ^{3/2} \epsilon  \left(1-2 \tilde{m} l_0^{-\frac{n}{2}}\right)+\frac{\tilde{m} \epsilon  l_0^{n-3} \Gamma (n)}{\Gamma \left(n-\frac{3}{2}\right)}}
\end{align}
\begin{align}
    K_2(0) = \frac{l_0^{n/2} \sqrt{\pi ^{3/2} \epsilon  \left(1-2 \tilde{m} l_0^{-\frac{n}{2}}\right)+\frac{\tilde{m} \epsilon  l_0^{n-3} \Gamma (n)}{\Gamma \left(n-\frac{3}{2}\right)}}}{\pi ^{3/4}}.
\end{align}
Consequently, these functions remain finite at the origin, which implies that $K_{\mu\nu}$ exhibits a discontinuity at $x = 0$, a hallmark of a thin-shell structure. Furthermore, we can evaluate the expansion scalar~\eqref{exp} for the two cases of interest, namely $n=2$ and $n=5/2$. In Fig.~\ref{expan}, we can clearly observe the recurrence of this discontinuity in $\Theta$ at the origin for both timelike and null geodesics. Note that for the case $k=0$ and $L=0$ in eq.~\eqref{expansion}, since $\Theta_0 = -\frac{4 E}{2r}$ is independent of $A$, the final result does not depend on $n$.

We can now evaluate the matter profile on the thin-shell layer that supports the bounce for the solution~\eqref{n22}. Substituting Eq.~\eqref{n22}  into Eqs.~\eqref{ener} and~\eqref{pre}, we obtain
\begin{eqnarray}
    \sigma = -\frac{ \sqrt{\pi ^{3/2} \epsilon  \left(1-2 \tilde{m} l_0^{-\frac{n}{2}}\right)+\frac{\tilde{m}  \epsilon  l_0^{n-3} \Gamma (n)}{\Gamma \left(n-\frac{3}{2}\right)}}}{2 \pi ^{7/4}l_0^{\frac{n}{2}} G},
\end{eqnarray}
\begin{eqnarray}
    P=\frac{\tilde{m} (\epsilon -1) l_0^{\frac{3 n}{2}} \Gamma (n)+\pi ^{3/2} l_0^3 \Gamma \left(n-\frac{3}{2}\right) \left(\epsilon  l_0^{n/2}-2 \tilde{m} \epsilon +\tilde{m}\right)}{4 \pi ^{7/4} G \Gamma \left(n-\frac{3}{2}\right) l_0^{n+3} \sqrt{\pi ^{3/2} \epsilon  \left(1-2 \tilde{m} l_0^{-\frac{n}{2}}\right)+\frac{\tilde{m} \epsilon  l_0^{n-3} \Gamma (n)}{\Gamma \left(n-\frac{3}{2}\right)}}}.
\end{eqnarray}
Consequently, we demonstrate that the energy density on the thin shell indeed violates the energy conditions. At this point, it is worth highlighting an important feature: the energy density described by the Lorentzian-like profile that sources the solution~\eqref{n22} does not violate the energy conditions everywhere, except at the bounce itself, where the metric~\eqref{n22} becomes degenerate. It is precisely at this location that we encounter a structure with a nontrivial topology described by the thin-shell formalism, which is ultimately sourced by a stress-energy tensor that violates the energy conditions.



Having established the true nature of the bounce for the solution~\eqref{n22}, we can now analyze the other hypersurface (should it exist) that cloaks this nontrivial topological structure, i.e., we shall investigate the event horizons. The event horizons of this line element can be found where $g_{tt}(x)=0$, so that
\begin{equation}
   x_h = \pm \sqrt{\left[\tilde{m}\pm\sqrt{\tilde{m} \left(\tilde{m}-\alpha_nl_0^{2n-3}\right)}\right]^{4/n}-l_0^2}.
\end{equation}
Note that this quantity is nonzero positive for $n > 3/2$. For
horizons to exist, i.e., we have a regular black hole where $\tilde{m}\geq \alpha_nl_0^{2n-3}$ and $l_0<\left[\tilde{m}\pm\sqrt{\tilde{m} \left(\tilde{m}-\alpha_nl_0^{2n-3}\right)}\right]^{2/n}$. For the extremal case, we have two possible conditions. The first occurs when 
$\tilde{m} = \alpha_n L^{\,2n-3}$ and $l_0^{\,3 - \frac{3n}{2}} \leq \alpha_n$, in which case the extremal horizon is located at $x_{\text{ext}} = \pm \sqrt{\tilde{m}^{\,\frac{4}{n}} - l_0^{\,2}}$. The second extremal horizon, located at the bounce, is obtained when 
$\tilde{m} \geq \alpha_n L^{\,2n-3}$ and $l_0^{\,\frac{n}{2}} = \tilde{m} \pm\sqrt{\tilde{m}\,(\tilde{m} - \alpha_n l_0^{\,2n-3})} $. Therefore, this latter case corresponds to the situation in which all horizons merge at the origin. For $\tilde{m} \leq \alpha_n l_0^{\,2n-3}$ we obtain a regular spacetime without horizons, described by a traversable wormhole. The same geometry can also be obtained when $\tilde{m} \geq \alpha_n l_0^{\,2n-3}$ and $l_0\geq\left[\tilde{m}\pm\sqrt{\tilde{m} \left(\tilde{m}-\alpha_nl_0^{2n-3}\right)}\right]^{2/n}$.

\begin{figure}[!ht] 
\centering
\includegraphics[height=5cm]{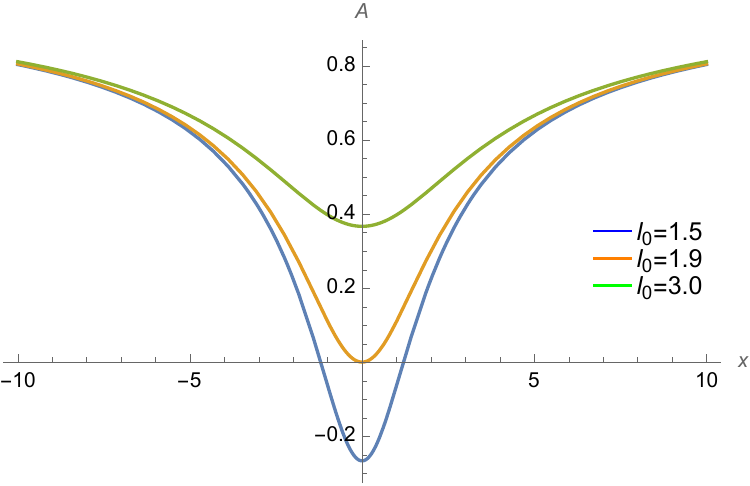}\quad
        \includegraphics[height=5cm]{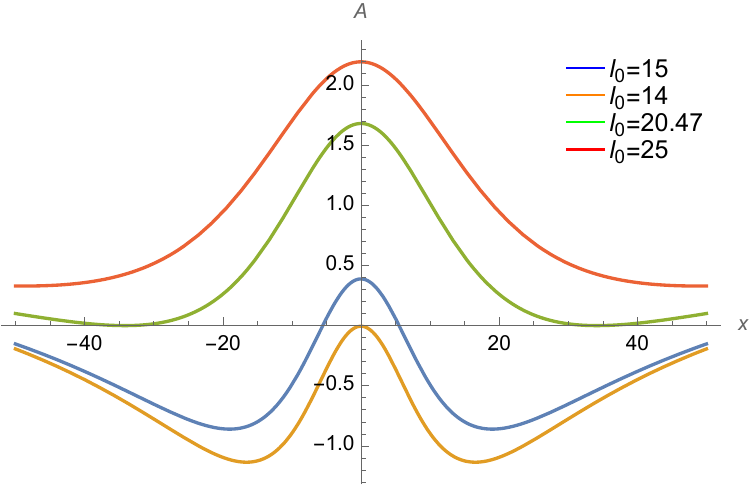}\quad
\caption{ The representation of the metric function $A$ from the line element (\ref{n22}) as a function of the radial coordinate $x$. The left panel corresponds to $n = 2$ with $\tilde{m} = 1$, while the right panel corresponds to $n = 5/2$ with $\tilde{m} = 100$. The bounce is located at the origin}  \label{fig1}
\end{figure}

To illustrate the possible compact objects, we consider two representative choices previously motivated by different quantum gravity theories, namely $n = 2$ and $n = 5/2$ in Fig.~\ref{fig1}. 
For the first case, we find a regular black hole solution with a single horizon (blue curve); moreover, we show that for $n = 2$ no solutions with more than one horizon can exist. We also obtain an extremal black hole located at the origin (orange curve), as well as a traversable wormhole (green curve).  In the case motivated by string T-duality, we find both regular black holes with a single horizon (orange curve) and regular black holes with two horizons (blue curve), consisting of an outer horizon and an inner (Cauchy) horizon. The extremal configuration is no longer located at the origin (green curve). The traversable wormhole is depicted by the red curve.  Therefore, these two cases exhibit markedly different behaviors.

\begin{figure}[!ht] 
\centering
\includegraphics[height=5.5cm]{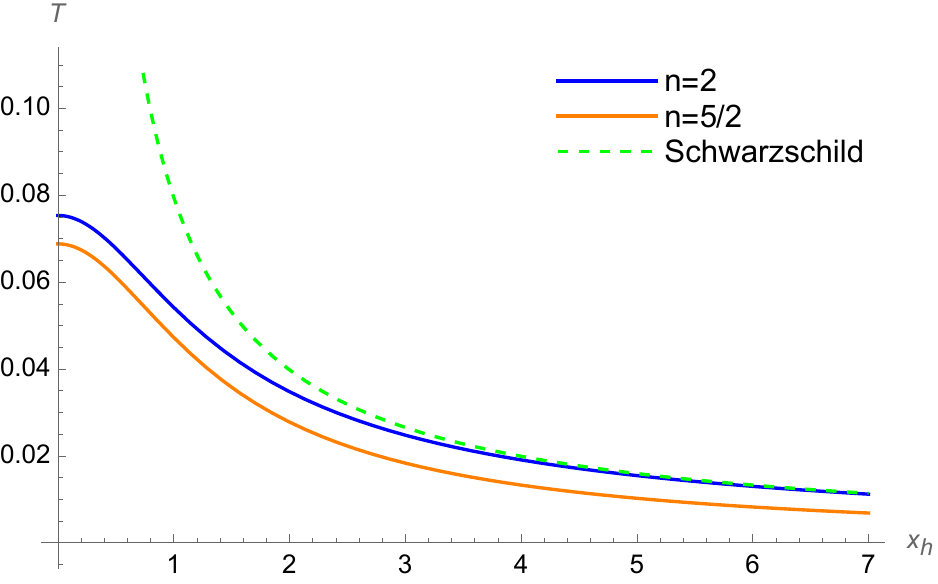}
\caption{ Hawking temperature as a function of $r_h$ of solution~\eqref{n22} assuming $l_0=1$.}  \label{t}
\end{figure}

It is now straightforward to compute the Hawking temperature of the solution~\eqref{n22} in the radial coordinate $x$, since $T_{H} = \frac{\kappa}{2\pi}$, where the surface gravity is given by
$\kappa = \frac{\partial_x A(x_h)}{2\sqrt{A(x_h)\,B(x_h)}}$. Thus, the Hawking temperature is
\begin{equation}
    T_{H} = \frac{1-\alpha_n  l_0^{2 n-3} \left(l_0^2+x_h^2\right)^{-\frac{n}{4}}}{4 \pi   \left(l_0^2+x_h^2\right)^{n/4}-2 \pi  \alpha_n  l_0^{2 n-3}}
\end{equation}
Note that for large black holes, i.e., $\frac{x_h^2}{l_0^2}>> 1$, and one recovers the standard result for the Hawking temperature: $T_{H} =\frac{1}{4\pi \tilde{x}_h}$, where once again it is necessary to perform the following transformation: $x_h^{n/2}=r_h$, where $r_h$ is the event horizon determined in the $r$ coordinate by setting $A=0$ in Eq.~\eqref{1mlorentz}. On the other hand, for small black holes it becomes clear that the effects of quantum–gravity corrections are significant and markedly different from the standard GR scenario; see Fig.~\ref{t}. During the evaporation process, the initial state shows that the temperature increases as the compact object evaporates. However, due to the quantum corrections associated with the minimal length $l_0$, the temperature no longer diverges as the process approaches its end. In fact, for $l_0 = 1$ and assuming the cases $n = 2$ and $n = \tfrac{5}{2}$, we find that the final stage of evaporation is characterized by a finite maximum temperature.

Another thermodynamic quantity of interest is the entropy. By definition the entropy as function of the ADM energy is $S = \int \frac{d\tilde{m}}{T(\tilde{m})}$. Thus, the entropy is given by
\begin{align} \nonumber
    &S = \frac{A}{4} + \frac{1}{2} \sqrt{\pi } \alpha_n l_0^{2 n-3}\sqrt{A} \\
    &+ \frac{1}{2} \pi  \alpha_n ^2 l_0^{4 n-6} \log \left[\pi ^{3/2} \Gamma \left(n-\frac{3}{2}\right) \left(\frac{\sqrt{A} l_0^3}{\sqrt{\pi }}-\alpha_n  l_0^{2 n}\right)\right]
\end{align}
where $A$ is the horizon area, given by $A = 4\pi \left(x_h^{\,2} + l_0^{\,2}\right)^{\frac{n}{2}}$.
Note that the logarithmic corrections to the entropy arise solely due to the quantum
nature of spacetime \cite{kaul2000logarithmic,carlip2000logarithmic}.

\subsection{Gaussian Distribution}
In addition to the Lorentzian-like profile~\eqref{dens}, another widely used choice in the literature, particularly within the framework of noncommutative gravity \cite{nicolini2006noncommutative,modesto2011black,rizzo2006noncommutative}, that exhibits a maximum energy density $\rho$ at the center is the Gaussian profile, which is given by
\begin{equation} \label{gauss}
    \rho(x) = \bar{M}e^{-\frac{x^2}{l_0^2}},
\end{equation}
where the parameter $l_0$\footnote{The characteristic length scale utilized here in the Gaussian profile does not necessarily have to coincide with the one employed in the profile~\eqref{dens}. However, for the sake of simplicity, we shall adopt the same notation for both, taking due care to distinguish between them whenever necessary.} is a length quantity defining the scale on which spacetime coordinates become quantum objects and $\bar{M}$ is the mass parameter. This means that for energies smaller than $1/l_0$ the function $\rho(x)$ approaches the Dirac delta distribution $\delta(x)$, i.e., the candidate quantum gravity is nothing but Einstein gravity. 
 Moreover, the total mass $M$ for this profile is  given by
\begin{equation}
   M = \int_{0}^{x} 4\pi \rho(x) x^2dx = \pi  l_0^2 \bar{M} \left[\sqrt{\pi } l_0\text{erf}\left(\frac{x}{l_0}\right)-2 x e^{-\frac{x^2}{l_0^2}}\right]
\end{equation}
where $\text{erf}(x)$ is the error function defined by $\text{erf}(x) = \frac{2}{\sqrt{\pi}}\int_{0}^{x}e^{-t^2}dt $. Again, assuming that $M = \tilde{m}$ in the limit $l_0 \to 0$ or $\frac{x}{l_0} \to \infty$ , we find that
\begin{equation}
    \bar{M} = \frac{\tilde{m}}{\pi^{3/2}l_0^{3}}.
\end{equation}

To obtain the areal function generated by this energy profile, we substitute Eq.~\eqref{gauss} into Eq.~\eqref{energ}, assuming once again that $\omega=1$, which yields
\begin{equation}\label{asyy}
    r(x)= r_0 e^{\frac{x^2}{4 l_0 ^2 }},
\end{equation}
where we set $\rho_0=\bar{M}$ and the size of the bounce at $x=0$ is $r(0)=r_0$.  Note, therefore, that unlike the previous case~\eqref{rlorentz}, the size of the bounce is not directly related to the minimal length, but rather to a free parameter $r_0$. Interestingly, near the origin the areal function can be expanded as
\begin{equation}
    r(x)\approx r_0\bigg(1 + \frac{x^2}{l_0^2}+\frac{x^4}{2l_0^4} + \ldots \bigg).
\end{equation}

As in the previous example, the expression for the metric function $A(x)$ is formally the same as in Eq.(\ref{eq:ASigma}) but with the parametrization given in Eq.(\ref{asyy}), namely, the line element for $\omega=1$ is given by

\begin{align} \nonumber \label{n2}
   & ds^2 = -\bigg(1-\frac{2\tilde{m}}{r_0}e^{-\frac{x^2}{4l_0^2}} +\frac{\tilde{m} r_0^2}{\pi ^{3/2}l_0^3}e^{-\frac{x^2}{2l_0^2}}\bigg)dt^2  \\
    &+ \frac{dx^2}{ \bigg( \frac{r_0^2 x^2 e^{\frac{x^2}{2 l_0^2}}}{4 l_0^4} \bigg) \bigg(1-\frac{2\tilde{m}}{r_0}e^{-\frac{x^2}{4l_0^2}} +\frac{\tilde{m} r_0^2}{\pi ^{3/2}l_0^3}e^{-\frac{x^2}{2l_0^2}}\bigg)}+  r_0^2 e^{\frac{x^2}{2l_0^2}} d\Omega^2.
\end{align}
Analyzing this solution in the radial coordinate $r$, we find that the metric component $A(r)$ is given by
\begin{equation}\label{as}
    A(r)=1-\frac{2\tilde m}{r}+\frac{\tilde{m} r_0^4 }{\pi^{3/2}l_0^3r^2} \ ,
\end{equation}
From Eq.~\eqref{asyy}, we conclude that the limit $\frac{x}{l_0} \to \infty$ is expressed in the  radial coordinate $r$ as $\frac{r}{r_0} \to \infty$, that is, $r_0 \to 0$. Therefore, in this limit the Schwarzschild solution is recovered in~\eqref{as}, as expected.

\begin{figure}[!ht] 
\centering
\includegraphics[height=5cm]{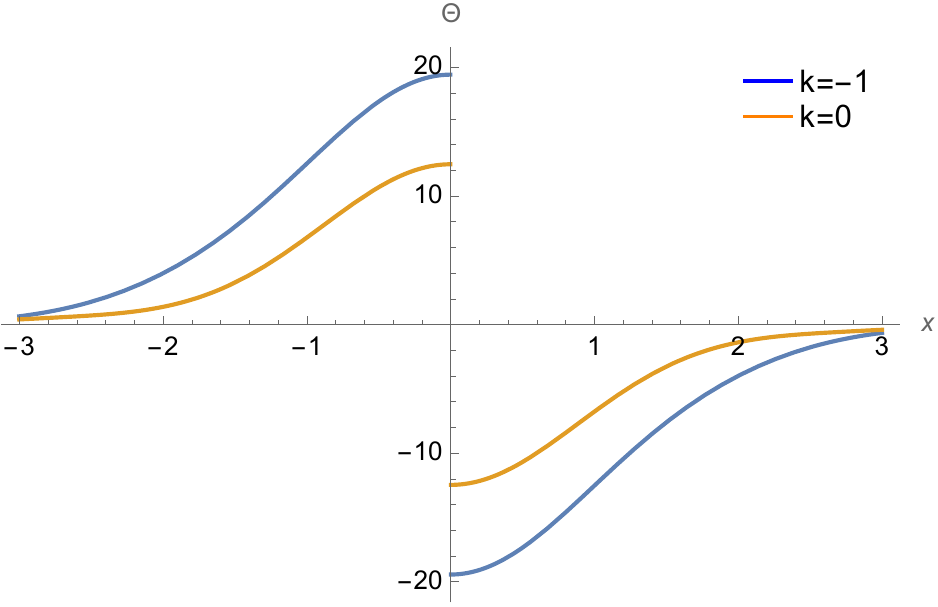}
\caption{ The representation of the expansion scalar $\Theta$~\eqref{exp} from the line element (\ref{n2}) as a function of the radial coordinate $x$ for $r_0=0.5$. We assume that $\tilde{m}=10$, $l_0=1$, $E=1$, and $L=1$. The bounce is located at the origin}  \label{expan2}
\end{figure}

Once the metric is found, we can calculate the components of the tensor $K_{\mu\nu}$. Substituting Eq.~\eqref{n2} into Eqs.~\eqref{k1} and~\eqref{k2}, we obtain
\begin{eqnarray}
    K_1(0) = \frac{\tilde{m} \left(r_0^5-\pi ^{3/2} l_0^3\right) \sqrt{\epsilon  \left(\frac{\tilde{m} r_0^4}{l_0^3}+\pi ^{3/2} \left(1-\frac{2 \tilde{m}}{r_0}\right)\right)}}{\pi ^{9/4} l_0^3 r_0^2},
\end{eqnarray}
\begin{equation}
    K_2(0) = r_0 \sqrt{\frac{\tilde{m} \epsilon  \left(\frac{r_0^5}{\pi ^{3/2} l_0^3}-2\right)}{r_0}+\epsilon }.
\end{equation}
Consequently, these components remain finite at the origin, implying that the extrinsic curvature tensor exhibits a discontinuity at this point due to the presence of the thin shell. This behavior is also directly reflected in the expansion scalar, as illustrated in Fig.~\ref{expan2}. Furthermore, this thin shell is supported by the following surface energy density:
\begin{eqnarray}
    \sigma = -\frac{\sqrt{\frac{\tilde{m} \epsilon  \left(\frac{r_0^5}{\pi ^{3/2} l_0^3}-2\right)}{r_0}+\epsilon }}{2 \pi  G r_0}
\end{eqnarray}
and
\begin{eqnarray}
    P = \frac{\pi ^{3/2} l_0^3 (-2 \tilde{m} \epsilon +\tilde{m}+r_0 \epsilon )+\tilde{m} r_0^5 (\epsilon -1)}{4 \pi ^{7/4} G l_0^3 r_0^2 \sqrt{\epsilon  \left(\frac{\tilde{m} r_0^4}{l_0^3}+\pi ^{3/2} \left(1-\frac{2 \tilde{m}}{r_0}\right)\right)}}.
\end{eqnarray}
Consequently, we confirm that the surface energy density on the thin shell exhibits a clear violation of the energy conditions. This reinforces the conclusion that both energy profiles, despite their structural differences, underpin qualitatively similar, non-trivial thin-shell topological geometries at the origin.

\begin{figure}[!ht] 
\centering
\includegraphics[height=5cm]{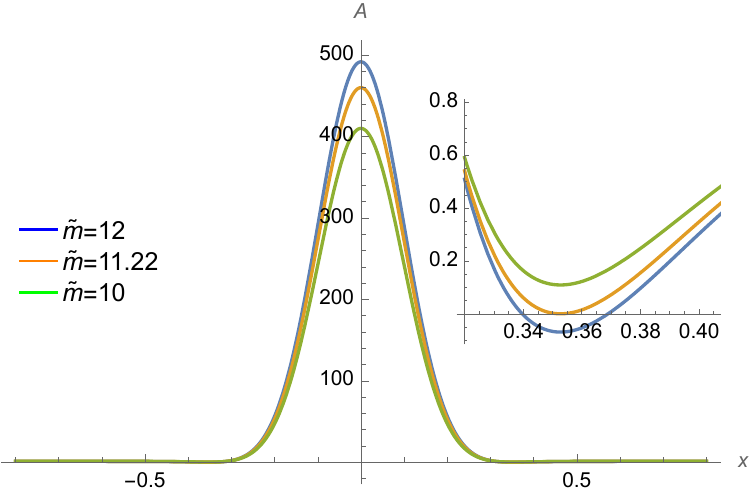}
\caption{ The representation of the metric function $A$ from the line element (\ref{n2}) as a function of the radial coordinate $x$ for $r_0=1$. The bounce is located at the origin}  \label{fig4}
\end{figure}

For completeness, we investigate the compact objects that can be described by the metric~\eqref{n2}, given the presence of a thin shell at the origin. To do so, we must examine the event horizons. The event horizons of the line element~\eqref{n2} can be found where $g_{tt}(x)=0$, yielding
\begin{equation}
   x_h = \pm 2 l_0 \sqrt{\log \bigg[ \frac{\tilde{m}}{r_0} \pm \frac{\sqrt{\tilde{m} \left(\pi ^{3/2} l_0^3 \tilde{m}-r_0^6\right)}}{\pi ^{3/4} l_0^{3/2}r_0} \bigg]}.
\end{equation}
In contrast to the Lorentzian-like case, in Fig.~\ref{fig4} we investigate the possible compact objects generated by the above metric by keeping the fundamental minimal length fixed at a small value (assuming $l_0 = 0.1$) while varying only the mass of the object. Once again, it was possible to find three classes of compact objects. The first is a black hole with two horizons (blue curve), followed by an extremal case (orange curve), and finally, a traversable thin-shell wormhole (green curve), all of which are asymptotically flat.

\begin{figure}[!ht] 
\centering
\includegraphics[height=5.5cm]{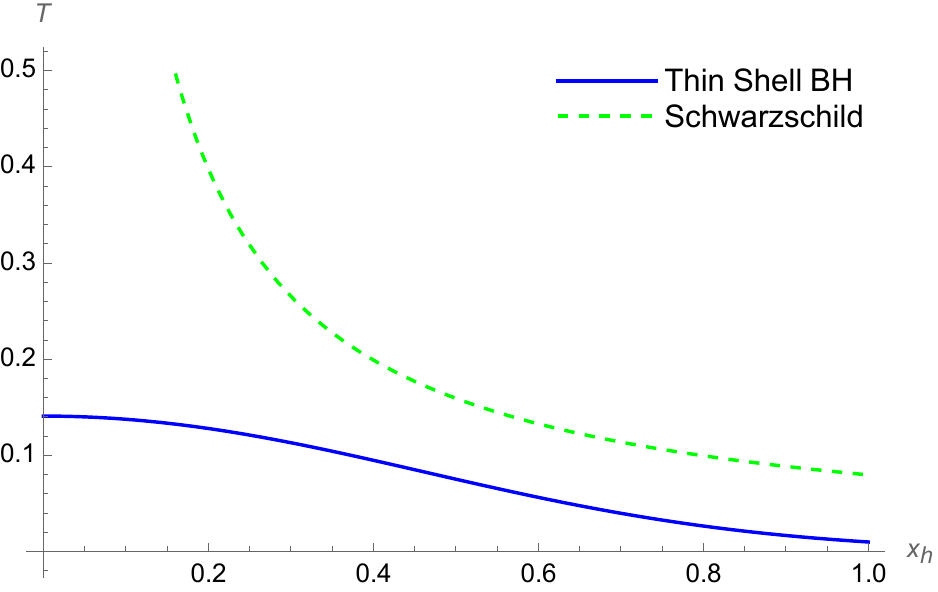}
\caption{ Hawking temperature as a function of $r_h$ of solution~\eqref{n2} assuming $l_0=0.3$ and $r_0=0.5$.}  \label{t2}
\end{figure}

Finally, let us analyze some thermodynamic properties of the solution~\eqref{n2}. Just as in the solution with a Lorentzian-type energy profile, the Hawking temperature for the Gaussian case is quite similar, as expected; that is, we observe the presence of a remnant in the final stage of the black hole evaporation process, as shown in Fig.~\ref{t2}. Furthermore, we also obtain the expression for the entropy, given by
\begin{equation}
 S =  \frac{A}{4} + \frac{\sqrt{A} r_0^6}{2 \pi  l_0^3}+\frac{r_0^{12} \log \left(r_0^5-\frac{\pi  \sqrt{A} l_0^3}{r_0}\right)}{2 \pi ^2 l_0^6},
\end{equation}
where $A$ is the horizon area, $A=4\pi r_0^2 e^{\frac{rh^2}{4l_0^2}}$. Note that we once again find a logarithmic correction arising from the quantum corrections of the model.

\section{Final Remarks}\label{con}
In this work, we have constructed a new class of static, spherically symmetric black bounce solutions. 
Working within an effective quantum-gravity framework, in which the point-like mass distribution of GR is replaced by a smeared energy density motivated by the existence of a fundamental minimal length, we have shown that a bounce necessarily arises when the matter, modeled by an anisotropic fluid,  satisfies a vacuum-like radial EoS, namely $\rho + p_r = 0$,  along with the tangential relation $p_t = \omega \rho$, provided the energy density exhibits a local maximum. This construction reproduces regular black hole and wormhole geometries, with the bounce structure emerging from the smeared profile.
 
Although the bulk anisotropic fluid satisfies the null, weak, strong, and dominant energy conditions for  $0\le \omega \le 1$, we have demonstrated that the metric is degenerate precisely at the bounce, where the inverse metric becomes ill-defined. A careful analysis of the extrinsic curvature and of the expansion scalar for geodesic congruences reveals a genuine discontinuity at this locus, signaling the presence of a hidden thin shell. Application of the Lanczos junction conditions shows that this shell is unavoidably threaded by a negative surface energy density, so that the energy conditions are necessarily violated locally at the bounce even though they hold everywhere else in the bulk. 

We have illustrated this general mechanism with two physically motivated smearing profiles, a Lorentzian-type distribution, associated with noncommutative geometry and string T-duality effects, depending on the choice of the exponent $n$, and a Gaussian distribution, characteristic of the coherent-state approach to noncommutative gravity. Both profiles yield asymptotically flat spacetimes that reduce to the Schwarzschild solution in the limit of vanishing minimal length, while producing markedly different near-origin structures: in the Lorentzian case the bounce radius is set directly by the minimal length scale $l_0$, whereas in the Gaussian case it is controlled by an independent free parameter $r_0$. Despite this difference, both cases admit the same three qualitative classes of compact objects: a regular black hole with one or two horizons, an extremal configuration, and a traversable wormhole, depending on the relation between the mass parameter and the minimal-length scale. It is worth highlighting that, although the solutions are characterized solely by the mass and quantum-minimum-length parameters, the metric features a modified Reissner-Nordström-like term that depends on these model parameters rather than a standard $U(1)$ gauge charge. This structure can lead to small deviations from general relativity predictions, even when the solution is expressed in the standard $r$-coordinates typically utilized for far-field localization~\eqref{1mlorentz}~\eqref{as}.


We investigate some implications of this solution for the thermodynamic analysis. In both the Lorentzian and Gaussian cases, the Hawking temperature deviates from the standard Schwarzschild behavior at small horizon radii and remains finite throughout the evaporation process, terminating in a finite-temperature remnant instead of diverging. The associated entropy receives explicit logarithmic corrections to the area law, which we demonstrate originate entirely from the quantum-gravity-motivated smearing of the source and vanish in the $l_0 \to 0$ limit.

Taken together, our results indicate that effective quantum-gravity corrections encoded through smeared matter distributions generically produce regular compact objects whose regularity is only achieved at the price of a thin-shell bounce violating the energy conditions locally. This suggests that the appearance of thin shells may be a rather generic feature of black-bounce constructions built from non-exotic, energy-condition-respecting bulk fluids, rather than a peculiarity of specific models. 

Natural extensions of this work include the study of the linear stability of these thin-shell bounces, the extension to rotating configurations, and the analysis of geodesic motion and observational signatures, such as shadows and quasinormal modes, that could in principle distinguish these thin-shell black bounces from their smooth counterparts. More broadly, we conjecture that the thin-shell bounce uncovered here may itself be understood as an effective, matter-supported realization of the degenerate-metric extensions of general relativity discussed in the topology-change literature, a connection we hope to make precise in future work.

\begin{acknowledgments}
The authors would like to acknowledge  Conselho Nacional de Desenvolvimento Cient\'ifico e Tecnol\'ogico (CNPq) from Brazil, for partial financial support. R. B . M is supported by CNPq/PDJ 151146/2025-0.
\end{acknowledgments}

\bibliography{biblio}


%
%
%
\end{document}